\documentclass[aps,prb,showpacs,preprint]{revtex4}

\usepackage{graphicx}
\usepackage{amsmath}
\usepackage{natbib}
\usepackage{bm}
\usepackage[dvips]{color}

\begin{document}

\title{
Structure optimization effects on the electronic properties of
Bi$_2$Sr$_2$CaCu$_2$O$_8$ }

\date{\today}
\author{V.~Bellini}
\email[Corresponding author: ]{bellini.valerio@unimore.it}
\affiliation{INFM-National Research Center on nanoStructures and
bioSystems at Surfaces (S3) \\ and Dipartimento di Fisica,
Universit\`a di Modena e Reggio Emilia, Via Campi 213/A, I-41100
Modena, Italy}
\author{F.~Manghi}
\affiliation{INFM-National Research Center on nanoStructures and
bioSystems at Surfaces (S3) \\ and Dipartimento di Fisica,
Universit\`a di Modena e Reggio Emilia, Via Campi 213/A, I-41100
Modena, Italy}
\author{T.~Thonhauser}
\affiliation{Department of Physics, The Pennsylvania State
University, University Park, PA 16802, USA}
\author{C.~Ambrosch-Draxl}
\affiliation{Institut f\"ur Theoretische Physik, Universit\"at
Graz, Universit\"atsplatz 5, A-8010 Graz, Austria}

\begin{abstract}
We present detailed first-principles calculations for the normal
state electronic properties of the high T$_C$ superconductor
Bi$_2$Sr$_2$CaCu$_2$O$_8$, by means of the linearized augmented
plane wave (LAPW) method within the framework of density
functional theory (DFT). As a first step, the body centered
tetragonal (BCT) cell has been adopted, and optimized regarding
its volume, $c/a$ ratio and internal atomic positions by total
energy and force minimizations. The full optimization of the BCT
cell leads to small but visible changes in the topology of the
Fermi surface, rounding the shape of CuO$_2$ barrels, and causing
both the BiO bands, responsible for the pockets near the
\textit{\=M} 2D symmetry point, to dip below the Fermi level. We
have then studied the influence of the distortions in the BiO
plane observed in nature by means of a $\sqrt{2}\times\sqrt{2}$
orthorhombic cell (AD-ORTH) with $Bbmb$ space group. Contrary to
what has been observed for the Bi-2201 compound, we find that for
Bi-2212 the distortion does not sensibly shift the BiO bands which
retain their metallic character. As a severe test for the
considered structures we present Raman-active phonon frequencies
($q = 0$) and eigenvectors calculated within the frozen-phonon
approximation. Focussing on the totally symmetric A$_{g}$ modes,
we observe that for a reliable attribution of the peaks observed
in Raman experiments, both $c$- and $a$-axis vibrations must be
taken into account, the latter being activated by the in-plane
orthorhombic distortion.
\end{abstract}
 \maketitle

\section{Introduction}
\label{intro}

The study of the electronic properties of the high-temperature
superconductors (HTSC) has been a major research topic in material
physics, since their discoveries in the late
eighties.\cite{bednorz1986} The normal metallic state ($T>$ T$_c$)
of such compounds has been itself a matter of controversy due to
the anomalous behavior of its resistivity and transport
properties, compared to classical metals. These anomalies seem to
point to a non Fermi liquid behavior of the normal state low lying
excitations, at least in the low doping regime, and to account for
it new theories have been necessarily postulated (see Ref.
\onlinecite{anderson1991} and references therein). It is equally
puzzling that many compounds belonging to different families of
HTSC cuprates with different chemical components and crystal
structures show on the other hand a remarkably similar behavior.
It is now generally accepted that the origin of these similarities
stems from the CuO$_{2}$ planar complex, existent in different
form in all the HTSC cuprate families, where the main physics
resides and where it is believed that the Cooper pairs form at low
temperature. The one-band Hubbard model, where only the strong
interactions in the CuO planes are considered, is able to
reproduce the flat quasiparticle band with Cu$_{\mathrm
d}$-O$_{\mathrm p}$ character which tends to pin the Fermi level
leading to a van Hove singularity near the X point (or
\textit{\=M} depending on the space group). Such a feature has
indeed been observed in angular-resolved photoemission (ARPES)
experiments.\cite{dessau1993}

Due to the refinement obtained by modern spectrometers, i.e.
better $\Delta E$ and $\Delta k$ resolution, very detailed
analysis of the Fermi surface topology started to appear in the
literature, renewing the interest in studying the electronic
properties in the normal and superconducting  state. Among the
HTSC cuprates, the two layer compound of the bismuth family, i.e.
Bi$_2$Sr$_2$CaCu$_2$O$_8$ (or Bi-2212 in a shorter notation), has
been chosen as a prototype for spectroscopic studies, due to the
relative ease of characterizing a stable crystalline surface not
exposed to segregation or oxygen depletion
phenomena.\cite{ori1995} A large number of studies can be found in
the literature (for a recent review see
Ref.~\onlinecite{damascelli2001} and reference therein), which
investigated the topology of the Fermi surface of Bi-2212 as a
function of photon energy, temperature and doping concentration.
There are two major complications in the analysis of the Fermi
surface topology. The first is the existence of matrix element
effects inherent in the photoexcitation process, which was
recently clarified by the theoretical work of Bansil, Lindroos,
and coworkers,\cite{bansil1999,lindroos2002} and supplied for a
unified picture on the long debated character, i.e. `hole' or
`electron'-like, of the Fermi surface. In addition, diffraction
experiments on the composition of Bi-2212 indicated the presence
of an incommensurate superstructure \cite{levin1994,
miles1998,etrillard2000} in the Bi-O plane (or Bi-O `modulation'),
near to a commensurate orthorhombic distorted cell $\sqrt{2}
\times 5\sqrt{2}$ with axes at $45^\circ$ with respect to the
in-plane Cu-O bonds. Refined ARPES measurements have shown the
existence of such `umklapp' bands produced by zone-folding along
the distortion direction, which overlap with the main Fermi sheets
at several points of the Brillouin zone.

In order to interpret the experimental data, detailed analysis
from first principles are therefore very much needed, which takes
into account all the chemical components of the material. The
first band structure calculations of Bi-2212
\cite{massidda1988,krakauer1988,hybertsen1988,szpunar1992}
appeared shortly after the discovery of the Bismuth family of
HTSC's. In all cases, the body centered tetragonal structure, or
the face centered structure has been adopted to describe the
crystal together with the experimental volume and atomic
positions; no real structure optimization has been attempted. Some
years ago Singh and Pickett \cite{singh1995} investigated the
influence of the orthorhombic distortion in the Bi-O plane on the
bands near the Fermi level, although they focused the attention on
the one CuO$_{2}$ plane compound, i.e. Bi-2201. This work clearly
shows that the distortion induces some variations in the binding
energy of the bands stemming from the BiO plane, and puts forward
that similar effects may play a role also in the two and three CuO
layers compounds, Bi-2212 and Bi-2223. The displacements from the
tetragonal positions have been indicated as a possible cause for
the absence of the BiO pockets at the Fermi surface predicted by
photoemission \cite{dessau1993} and, more unambiguously,
tunnelling experiments.\cite{tanaka1989,shih1991}

In the following we present a refined analysis of the electronic
properties of Bi-2212 by means of first-principles density
functional techniques. The work is organized as follows. After a
brief introduction of the Bi-2212 crystal structure in
Sec.~\ref{structure}, the method in use and some computational
details are presented in Sec.~\ref{method}. The discussion of the
results will be addressed in two separate sections: In
Sec.~\ref{bct0} we will still rely on the BCT cell, optimizing
cell parameters and internal atomic positions, while in
Sec.~\ref{orth0} a possible improvement over the tetragonal
description in terms of an ``averagely-distorted'' orthorhombic
(AD-ORTH) cell will be presented. The main focus will be on the
Fermi surface topology, and generally, on the band structure close
to the Fermi level, and, for the first time to our knowledge, on
the calculations of the Raman-active phonon frequencies and
eigenvalues from first principles.

\section{Crystal structure: BCT {\it vs.} AD-ORTH}
\label{structure}

Bi-2212 belongs, together with the one layer (Bi-2201) and
three layers (Bi-2223) compounds, to the bismuth HTSC cuprate
family. The basic unit is composed of two CuO$_{2}$ planes per
cell (one Cu and two O atoms per plane) separated by Ca ions, and
two Bi-O layers separated from each of the CuO complexes by a Sr-O
layer. The Bi-O planes have, similarly to the Cu-O chains in the
Yttrium family, the role of a `charge reservoir', attracting
electrons and therefore doping the CuO$_{2}$ planes with holes.
When the complex superstructures in the BiO plane observed by
diffraction experiments are neglected, the crystal structure of
Bi-2212 is well represented by the body centered tetragonal  cell
(with space group \textit{I4/mmm}) shown in
Fig.~\ref{fig1}(a). The superstructure arises because of the
mismatch between the equilibrium Bi-O bond length and the lattice
constant imposed by the CuO$_{2}$ planar nets. The weak coupling
of the two BiO planes has been found to favor the appearance of
such distortion, as opposed to the case of the Tl-2212 compound,
where the shorter distance between the TlO planes is
responsible for its absence. Moreover, the observed superstructure
is incommensurate with respect to the Cu-O lattice constant, and
is only approximately arranged by a $\sqrt{2}\times 5\sqrt{2}$
orthorhombic cell (see for instance Fig.~1 in Ref.
\onlinecite{levin1994}). The full account of these distortions
represents a severe task even for modern computers, as far as
first-principles supercell calculations are involved. We therefore
chose a more simplistic approach, considering the average
displacements of the atoms from their tetragonal sites. Insets (b)
and (c) in Fig.~\ref{fig1} show how the displacements of the
O$_{Bi}$ and Bi atoms in the BiO planes, allowed by an
orthorhombic $\sqrt{2}\times \sqrt{2}$ cell (with space group
\textit{Bbmb}), account for the mean expansion and contraction of
the Bi-O bonds. The type of distortion is the same as simulated by
Singh and Pickett \cite{singh1995} for the one layer Bi-2201
compound, whereas we stress that for the Bi-2201 crystal the
$\sqrt{2}\times \sqrt{2}$ cell itself is the commensurate analogue
to the real incommensurate superstructure. Here, it suffices to say
that in-plane average displacements of around $0.5$ \AA\ ($0.14$ \AA)
along the distortion direction are observed for the O$_{Bi}$ (Bi,
O$_{Sr}$) atoms.\cite{miles1998} More details on the experimental
lattice parameters as well as planar and off-plane coordinates of
all the atoms in the BCT and AD-ORTH cells will be given in tables
later in the paper.

\section{Method and computational details}
\label{method}

All calculations have been carried out using the full-potential
linearized augmented plane-wave (FP-LAPW) method
\cite{andersen1975,singh1994b} and its recent extension (APW+lo)
as implemented in the WIEN2k code.\cite{wien2k} The muffin-tin
radii inside which the plane waves are augmented by radial
functions expanded over spherical harmonics have been chosen to be
1.9 (Ca), 1.9 (Cu), 2.2 (Sr), 2.25 (Bi), and 1.45 (O) Bohr.
Exchange and correlation effects are accounted for by the local
density approximation (LDA). In the wave function expansion
$\approx$ 2200 (2400) basis functions at the $\Gamma$ point have
been used, while 40 (27) special $\bf{k}$ points within the
irreducible part of the BZ sufficed for Brillouin-zone (BZ)
integrations for the BCT and AD-ORTH structures, respectively.
Starting from the experimental parameters we have optimized the
volume and $c/a$ ratio for the BCT (and also $b/a$ for the AD-ORTH
cell) by minimization of the total energy. At each step the atoms
were allowed to relax to their equilibrium positions under the
influence of the Hellmann-Feynman forces, a  procedure leading to
very accurate equilibrium structures. The remanent forces in this
case are less than $0.2$ mRy/a.u., which is an important
ingredient for the estimation of reliable phonon frequencies. A
detailed description of the atomic force calculation within the
LAPW method is given in Ref.~\onlinecite{kouba1997}. Within the
frozen-phonon approach a polynomial fit of calculated atomic-force
values is carried out. For the fully symmetric $A_{g}$ modes this
was done for the equilibrium position plus two to four different
displacements of each participating atom along the Cartesian axis.
Diagonalisation of the dynamical matrix yields the phonon
frequencies as well as the normal vectors of the vibrations.

\section{Results for the BCT cell}
\label{bct0}

\subsection{Structural optimization}
\label{bct1}

Band structure calculations adopting tetragonal (or almost
tetragonal) body centered or face centered cells for the
description of the Bi-2212 crystal go back to the late
eighties,\cite{massidda1988,hybertsen1988,krakauer1988,szunyogh1991}
shortly after the discovery of high temperature superconductivity
in doped La$_2$CuO$_4$\cite{bednorz1986} and other cuprates. In
all cases lattice parameters taken from experiments have been used
and no optimization of the volume and atomic coordinates has been
attempted. It has been instead demonstrated recently during the
study of another HTSC, \cite{kouba1999} i.e. YBa$_2$Cu$_3$O$_7$,
that the optimization of the crystal parameters (volume, $c/a$
ratio, atomic coordinates) within the local density approximation
lead to a better agreement with experiments for what concerns the
Fermi surface topology and the values of Raman-active phonon
frequencies. Starting from the experimental data by Sunshine et
al. \cite{sunshine1988} (Exp) we have optimized the volume
(Vol-Opt) or both, volume and $c/a$ ratio (Full-Opt) by standard
total energies techniques (see Table~\ref{tab1}). Atoms were
allowed to relax to their equilibrium positions at each step of
the minimization procedures. The case where internal coordinates
are relaxed using the experimental lattice volume and $c/a$ ratio
(Atom-Opt)  is also given. As evident from Table~\ref{tab1} the
well known LDA underestimation of the volume amounts to more than
$10$\% (theory $202.2$ \AA$^3$, experiment $226.3$ \AA$^3$), while
for YBCO \cite{kouba1999} it was around $6 $\% only. The optimized
$c/a$ ratio is $2$\% larger than the experimental one, but its
influence on the cohesive energy and the equilibrium positions of
the nuclei is small. The band structure for the experimental cell
agrees very well with other LAPW
results,\cite{massidda1988,krakauer1988} and will not be given
here. Both, the volume reduction and the increase in the $c/a$
ratio results in a squeezing of the Cu-O in-plane bond length; as
a net effect, as will be evidenced in the next section,
hybridization between CuO$_2$ and BiO planes through the SrO plane
is favored inducing both BiO bands to cross the Fermi level. This
is consistent with the results obtained by Szpunar et al.\ in
Ref.~\onlinecite{szpunar1992}, who investigated the effect of
stress along the $c$ axis by decreasing the c/a ratio up to
$12$\%, and found consistently the opposite trend, i.e. the BiO
bands move to lower energies. For what concerns the atomic
coordinates within the cell, a relevant change is the sizable
increase of the dimpling in the SrO plane (of around $0.3$ \AA),
while it is only slightly reduced in the CuO$_2$ and BiO planes.
The dimpling of the CuO$_2$ plane in HTSC has been shown to be
connected to the bifurcation of the saddle point near the
\textit{\=M} point,\cite{andersen1994} and much attention has been
devoted to it in the framework of the van Hove
scenario.\cite{markiewicz1997}

\subsection{Fermi surface}
\label{bct2}

Fig.~\ref{fig2} depicts the calculated cross cuts of the Fermi
surface with the $k_z=0$ plane. The dispersion along the direction
perpendicular to the plane is very low due to the high 2D
character of the layered cuprates. The contour plots have been
obtained by associating a sharp Lorentzian to each eigenvalue and
summing the contribution at the Fermi level from all the bands at
each k-point; the small broadening induced in this way has
therefore no physical origin. The plots are centered around the
$X\equiv(\pi/a,\pi/a)$ point, while $\Gamma\equiv(0,0)$ and
$Z\equiv(0,2\pi/a)$ (or equivalently $(0,\pi/a)$) are at the
corners of the squares. Moving from the experimental structure
(Fig.~\ref{fig2}(a)) to the fully optimized structure
(Fig.~\ref{fig2}(b)) we observe small but visible changes in the
Fermi surface topology. The two main barrels centered at $X$,
arising from the Cu$_{d}$-O$_{p}$ bands, giving rise to the well
known hole-like Fermi surface, attain a more rounded shape and the
splitting (usually indicated as ``bilayer splitting'') of the
bonding and antibonding parts is generally reduced. Near the
\textit{\=M} point, where the splitting is expected to be larger,
the interpretation is complicated by the hybridization with the
BiO orbitals, which induces an anticrossing between the bands.
Both BiO bands, with antibonding character and $p_{(x,y)}$
symmetry, are found to cross the Fermi level for the optimized
structure, giving rise to two distinct intersecting pockets around
\textit{\=M}, which dope the CuO$_2$ planes with additional holes.
The main issue about these pockets is that there is no evidence of
them in both, photoemission and tunnelling experiments, which is
normally attributed to the incommensurate structural
modulation/distortion present in the BiO plane. (We will come back
to this issue in Sec. \ref{orth2}.) Bansil, Lindroos and coworkers
\cite{bansil1999,lindroos2002} have shown by means of the KKR
method and one-step photoemission theory that matrix elements
modulate the intensity of the spectra at the Fermi level
differently depending on the wave vector and energy of the
incoming photon. Yet, in all their calculations, the potentials
were adjusted in an ad-hoc manner in order to shift the BiO bands
above the Fermi level. While matrix elements effect and low
scattering cross-sections for the bands in the BiO plane might
explain their absence in ARPES measurements, tunnelling experiment
on the other hand are less criticizable. Notwithstanding, it has
been shown recently that filtering effects stemming from the
symmetry of the surface' electronic states are responsible for the
observed suppression of the tunnelling conductance in case of a
CuO terminated surface\cite{misra2002}, and therefore a simulation
of scanning tunnelling spectra on BiO-plane terminated surfaces
might help in the future to clarify this argument. The explicit
treatment of correlation effects in the CuO$_2$ plane beyond the
mean-field does not modify this picture leaving the hybridization
between Cu and BiO bands unaltered. Work along this line is under
way and will be presented in a separate
contribution.\cite{tobepublished1}

Since photoemission and even more tunnelling experiments probe
only the surface layers of a material and since the BiO plane is
the natural cleavage plane during sample preparation, one might
argue that the surface BiO bands might slightly deviate from the
bulk ones. We have therefore investigated this possibility and
simulated the presence of a BiO plane terminated (001) surface.
Thereby we used a supercell technique which considers repeated
slabs, each composed of two halves of BCT cells embedded in
vacuum. The in-plane lattice constant was the one determined in
the optimization of the bulk, while the atoms were allowed to
relax along the direction perpendicular to the surface. Such
relaxations were found to be of limited size, consistent with the
highly two-dimensional nature of the compound, but still induced
slight changes in the Fermi surface (see Fig.~\ref{fig2}(c)). The
dimpling in the BiO plane doubles in value (from 0.09 to 0.16 \AA)
and the BiO bands becomes almost degenerate crossing the
$\Gamma$-\textit{\=M} direction at the same point. The CuO barrels
move towards the \textit{\=M} point and the BiO pockets becomes
less distinguishable. Overall, the so optimized Fermi surface
compares better with ARPES measurement (see for instance Ref.
\onlinecite{aebi1994,bogdanov2001,sato2001a,feng2002b,kordyuk2002,asensio2003}).

\subsection{Raman-active phonons}
\label{bct3}

A rigorous test for the quality of any structural
characterization, is the calculation of the phonon frequencies,
which, through the dynamical matrix, gives an account on how well
the equilibrium positions as well as the bonding between the atoms
are described. For the tetragonal cell, 14 Raman active ($q=0$)
modes (6$A_{1g}$ + 1$B_{1g}$ + 7$E_{g}$) are predicted by the
\textit{I4/mmm} space group. The $A_{1g}$ modes are symmetric
\textit{c}-axis vibration of Bi, Sr, Cu, O$_{Bi}$, O$_{Sr}$ and
O$_{Cu}$ atoms, the $B_{1g}$ phonon is the out-of-phase motion of
the O$_{Cu}$ atoms also along the \textit{c}-axis, and E$_{g}$
vibrations are planar \textit{ab} displacements. The frequencies
of the $A_{1g}$ c-axis Raman-active modes and the corresponding
eigenvectors, obtained by the frozen phonon approach, are listed
in Table~\ref{tab2}. We start the discussion by saying that while
the frequencies of the measured phonon peaks agree well under
similar experimental conditions, their assignments have changed
sensibly during the last decade and among different groups.
Focusing only on the most recent investigations,\cite{note1} the
strongest dispute concerns the origin of the two highest
frequencies, i.e. at around 460 and 630 cm$^{-1}$, which were
attributed to the motion of the O$_{Bi}$ and O$_{Sr}$ atoms
\cite{kakihana1996,chen1998,osada1999,williams2000,pantoja1998},
respectively, or viceversa.\cite{kendziora1997} The frequency
around 115 and 130-140 cm$^{-1}$ is concordantly believed to stem
from Sr and Cu vibrations along the c-axis. On the other hand,
fewer investigations focussed onto the low frequency part of the
spectrum where the phonons of the heavier atom, i.e. Bi, should
appear, and a well resolved peak at around 60 cm$^{-1}$ has been
successfully observed.\cite{osada1998,boekholt1992,sapriel1991}
Attribution of the peaks in the Raman spectra have been achieved
either by investigating the change in frequencies and/or
intensities upon substitution with Y or Pb atoms, or by
isotopes,\cite{pantoja1998} or by comparing with existent
theoretical calculations. For the former type of analysis, the
attribution strongly depends on the choice of the atomic site
where the substitution is believed to take place. From the
theoretical side, lattice dynamics calculations in the framework
of the shell model have been performed more than a decade ago
\cite{prade1989} and found the 6 A$_{1g}$ phonons to be placed at
87, 164, 182, 387, 493 and 517 cm$^{-1}$. A shortcoming of this
method is that the choice of the interatomic potentials, which
enter the calculations as parameters, is not unique and
consequently interactions in the crystal might be not accounted
for properly.

As can be inferred from columns 2-7 of Table~\ref{tab2}, which
list the normal vectors of phonons, we attribute the 6
frequencies, in increasing order of eigenvalue, to the main
vibrations of Bi, Cu, Sr, O$_{Bi}$, O$_{Cu}$ and O$_{Sr}$. While
the motion of the three oxygens have a somewhat purer character,
the lower vibrations involve the in- and out-of-phase motion of
more than one atom. The most striking deviations from the
experiments are found for the vibration of the oxygen atom in the
BiO plane (233 vs. 460 cm$^{-1}$), and the opposite assignment of
the phonons at 116 and 163 cm$^{-1}$. Our calculations support
those experiments that attribute the frequency at around 630
cm$^{-1}$ to the c-axis vibration of the apical O$_{Sr}$ atom.
Reasonable but not satisfactory agreement is found therefore only
for the three phonons stemming from the Bi, O$_{Cu}$ and O$_{Sr}$
atoms, which are found to be around 15\% weaker than their
experimental counterparts.

What might be the reason for this disagreement? Regarding the
opposite attribution of the 116 and 163 cm$^{-1}$ phonons, we
might say that explicit inclusion of correlation effects beyond
the mean field in the Cu \textit{d} bands are expected to change
the bond strength and orbital localization in the CuO$_2$ plane,
and consequently influence the restoring force acting on the Cu
and O atoms. On the other hand LDA (or GGA)-frozen phonon
calculations have demonstrated in the recent past to be able to
reproduce very accurately the measured Raman-active phonons of
other correlated superconductors such as YBaCuO \cite{kouba1999},
La$_{2-x}$Ba$_x$CuO$_4$ \cite{thonhauser2003} and
\textit{R}Ni$_2$B$_2$C (with R=Y,Lu) \cite{ravindran2003}, and it
is therefore puzzling why only for Bi-2212 correlation effects
should play a role. For what concerns the vibration of the
O$_{Bi}$ atom, the calculated vibrational frequency is only half
of the measured value, a deviation that is too large to be ascribed
to the approximation adopted in the method. The inadequacy of the
tetragonal approximation for the structural characterization of
Bi-2212 compound seems therefore the only plausible explanation
for this large discrepancy. Expansion and contraction of the Bi-O
bonds along the distortion direction, induced by the displacement
of the atoms from the tetragonal positions, might change the
nature of the bonds together with the mutual distances between the
atoms and consequently the stiffness of the restoring force.
Moreover modes that are forbidden in a tetragonal cell become
Raman-active in an orthorhombic one, and for large atomic
displacements even the $q=0$ selection rules may be affected and
additional modes may appear. We will investigate this point
in Sec.~\ref{orth3}.

\section{Results for the AD-ORTH cell}
\label{orth0}

\subsection{Structural optimization}
\label{orth1}

Similarly to what has been done for the BCT cell, we present in
Table \ref{tab3}, a comparative listing of structural data for the
AD-ORTH structure. Now both, the $c/a$ and $b/a$ lattice constant
ratios have been optimized in the AD-ORTH structure. In
Table~\ref{tab3} the labels 'Full-Opt', 'Vol-Opt', 'Exp-Opt' have
the same meaning as in Table~\ref{tab1}. Experimental data for the
$Bbmb$ cell have been taken from Ref. \onlinecite{miles1998}. In
addition to the $z$ coordinates of the atoms, the $x$ coordinates
are given; the origin of the cell is shifted by $(a/2,b/2,-c/4)$
from the one of the BCT cell, while the orthorhombic $x$ and $y$
axis are rotated by $45^{\circ}$ with respect to the tetragonal
cell and point along the Bi-O bond directions. Compared to the BCT
cell, a smaller reduction of the volume is found by LDA (4.2\% vs.
10\%); the $c/a$ ($b/a$) ratio increases (decreases) by about 2
(3.4)\%. The atomic $z$-coordinates are generally very similar to
the ones in the BCT cell, while only a slight enhancement of the
dimpling in SrO plane is found. As depicted in Fig.~\ref{fig1},
the average displacements from the tetragonal positions of the Bi
and O atoms in the BiO layers induce alternate expanded and
contracted Bi-O bond distances along the $x$ direction, which
amount to 2.69 $\pm$0.64 \AA\ and to 2.67 $\pm$0.55 \AA for the
experimental and fully optimized structure, respectively. Along
the other bond direction, which is now distorted into a zig-zag
line, the average bond distances are 2.72 \AA\ and 2.59 \AA for
the experimental and optimized cell, respectively. Total energy
calculations performed with the experimental lattice constants
show that the distorted structure is more stable than the
tetragonal one; the difference in the total energy amounts to
0.60~eV per formula unit, similarly to what has been found for the
Bi-2201 compound.\cite{singh1995} This is a first clear indication
that the tetragonal description is far from being realistic, and,
as put forward for the Bi-2201 compound, one should not entirely
rely on it when comparing with ARPES or Raman experiments.

\subsection{Band structure and Fermi surface}
\label{orth2}

The large structural distortion in the BiO plane of the one layer
bismuth cuprate, i.e. Bi-2201, has been
demonstrated\cite{singh1995} to induce a sensible (400 meV)
upwards shift of the partially occupied BiO bands that dip below
the Fermi level near the \textit{\=M} point of the tetragonal
Brillouin zone. We plot in Fig.~\ref{fig3} the band structure  in
the vicinity of the Fermi level for Bi-2212 in the BCT (a) and
AD-ORTH structure (b), along the $\Gamma-M$ line of the
orthorhombic Brillouin zone; The same k-point direction has been
chosen for Bi-2201 so that Fig.~\ref{fig3} and Fig.~3 in Ref.
\onlinecite{singh1995} are directly comparable. The \textit{\=M}
2D symmetry point of the tetragonal cell sits in the middle of the
$\Gamma-M$ line. Four of the six bands which cross the Fermi level
stem from the BiO plane (doubled by the folding) while the other
two are the bilayer-splitted CuO bands mirrored across the zone
boundaries due to the $X\rightarrow\Gamma$ folding. In spite of
the similarities between the one and two CuO layers compounds
discussed above, it can be inferred from Fig.~\ref{fig3} that the
distortion in the Bi-2212 does not sensibly move the BiO bands
neither it visibly reduces the hybridization with the CuO bands
below. The different behavior of Bi-2201 and Bi-2212 might be
ascribed to the stronger interactions between the BiO and CuO$_2$
planes found in latter; the larger overlap between the orbitals is
also testified by the larger number of hole-carriers per CuO$_2$
plane transferred by the BiO plane.

As a result, now there are many more bands that cross the Fermi
level enormously complicating the topology of the Fermi surface,
as shown in Fig. \ref{fig4}. While multiple CuO barrels have been
detected in several experiments and correctly assigned to the
scattering of the photoemitted electron from the orthorhombic
distorted BiO planes, no clear evidence of neither one nor several
BiO pockets have been found. The similarities between the Bi-2201
and Bi-2212 Fermi surfaces, evidenced by ARPES experiments, on the
other hand suggest that the broadening of the bands due to the
incommensurate and disordered nature of the distortion together
with finite energy resolution of the spectrometers might justify
why many of the details of the Fermi surface are smeared out and
do not result into clear features at the Fermi level. Moreover,
the photoemission cross section from electrons in the BiO plane is
also much smaller than for the CuO$_2$ planes, which might
compromise their detection at the Fermi level.

\subsection{Raman-active phonons}
\label{orth3}

In most of the Raman scattering experiments present in the
literature the assignment of the phonon modes have been carried
out starting from the \textit{I4/mmm} space group used for the
tetragonal cell. But, since more phonon peaks are usually observed
in the spectra than what is expected in a tetragonal structure, it
is also a common opinion that several other vibrations become
Raman active due to the distortion or to folding at the Brillouin
zone edge. As for the BCT cell, also for the AD-ORTH cell the Ca
atoms are placed at the centers of inversions and associated
vibrations are therefore Raman inactive. A group-theoretical
analysis within the \textit{Bbmb} space group of the orthorhombic
cell \cite{sugai1989,liu1992} already appeared in the literature.
Limiting ourself to the study of the Raman-active phonons, each
set of Cu, Sr, Bi, O$_{Sr}$ and O$_{Bi}$ atoms in the cell (with
$m$ point group) contributes with 2 A$_g$ + 1 B$_{1g}$ + 1
B$_{2g}$ + 2 B$_{3g}$ modes. The two inequivalent sets of O$_{Cu}$
atoms in the CuO$_2$ plane, having different point groups, i.e.
$2$, contribute each with 1 A$_g$ + 1 B$_{1g}$ + 2 B$_{2g}$ + 2
B$_{3g}$ modes. Focussing only on the vibrations which do not
lower the symmetry of the crystal, i.e. A$_g$, we expect then to
find 12 A$_g$ modes, which in a first approximation separate into
7 vibrations along the c-axis and 5 along the a-axis. Among the
c-axis vibrations, only one of them was not active in the
tetragonal cell, and stems from the out-of-phase vibration of the
two (now inequivalent) O$_{Cu}$ atoms in the CuO$_2$ plane. Note
that the out-of-phase vibration of the O$_{Cu}$ ions is much lower
in frequency compared to its counterpart in YBCO. The reason for
this should be ascribed to the fact that different displacement
patterns are involved due to different point symmetries of the
ions in the unit cells.

In Table~\ref{tab4} the frequencies of all the 12 A$_g$ modes are
listed together with the associated eigenvalues. The first thing
we notice is that due to the orthorhombic distortion, the $c$-axis
vibration of the O$_{Bi}$ atom strongly couples with the $a$-axis
vibration of the apical O$_{Sr}$ atoms. As a net results the two
frequencies appear at 289 and 192 cm$^{-1}$. Both modes involve
also the vibration of Sr along the $a$-axis. It might be described
as a breathing mode in the SrO plane where an enhancement of the
orthorhombic distortion arises whenever the O$_{Bi}$ atom move
towards the SrO plane. Only 5 pure in-phase $c$-axis vibration
remains at 597, 367, 163, 104, 47; they all compare closely to the
one obtained for the tetragonal cell (see Table~\ref{tab2}). The
two highest modes at 597 and 367 have increased in the range of
10-20 cm$^{-1}$, while the Cu vibration decreases by the same
amount. The vibrational frequencies associated to Bi and Sr are
barely affected. The out-of-phase vibration of the oxygen atoms in
the CuO$_2$ plane has a frequency of 210 cm$^{-1}$. The O$_{Bi}$
$c$-axis vibrational frequency has moved from 233 cm$^{-1}$ for
the BCT cell to 289 cm$^{-1}$ but remains quite far from its
experimentally attributed value of 460 cm$^{-1}$.

In addition to the modes with pure or mixed c-axis character
discussed above there are 4 pure $a$-axis vibrations, as shown in
Table~\ref{tab4}. $a$-axis vibrations of the Bi and Sr atoms are
found at 106 and 132 cm$^{-1}$, respectively. Vibrations of the Cu
atoms instead results to be much harder, at a frequency of 355
cm$^{-1}$, three times larger than the corresponding vibration
along the $c$-axis. This large value should not completely
surprise considering that it stems from a vibration which compress
and expand the Cu-Cu distance along the bond axis. The most
interesting finding is the O$_{Bi}$ $a$-axis mode, which attains a
frequency of 424 cm$^{-1}$, almost twice the value calculated for
its $c$-axis vibration in the tetragonal cell. It is important to
recall that all these modes have A$_g$ symmetry and are thus
expected to appear in Raman spectra taken under non-crossed
polarization geometry such as (zz), (xx) or (yy). In light of
these results, it is therefore very tempting to associate the
experimental peak around 460 cm$^{-1}$ to the $a$-axis rather than
the $c$-axis vibration of the O$_{Bi}$ atom. Calculations of Raman
spectra under different scattering geometries might be able to
address this important issue in the future and allow for a direct
comparison with the experimental spectra.

\section{Conclusions}
\label{conclusion}

We have performed extensive  first-principles calculations for the
electronic structure of the two layer bismuth cuprate
superconductor Bi-2212. By means of the APW+lo method we have
optimized the crystal structure for both, a body centered
tetragonal cell as well as by taking into account the average
distortions in the BiO planes by an orthorhombic $\sqrt{2} \times
\sqrt{2}$ cell. When the full optimization of the BCT cell is
performed, the main two CuO barrels which characterize the Fermi
surface attain a more rounded shape and compare better with what
has been measured by ARPES experiments. Calculations show how both
the BiO bands at the \textit{\=M} symmetry point of the Brillouin
zone cross the Fermi level for the optimized structure. The
results obtained for a BiO terminated (001) surface exhibit some
changes in the details of the Fermi surface, yet of small size due
to the two-dimensional layered nature of the double perovskite
structure. An analysis of the band structure along the $\Gamma-M$
symmetry line of the orthorhombic AD-ORTH cell, shows that the
distortion, despite its large size, leaves the metallic character
of the BiO plane unmodified and does not sensibly change the
hybridization between the BiO and CuO planes. These findings
differ from what has been observed by Singh et al.\cite{singh1995}
in the one-layer Bi-2201 compound. In agreement with that work, we
predict the distorted structure to be more favorable than the
tetragonal one by $0.6$~eV per formula unit.

We have then presented first-principles calculations of Raman
active phonons within the frozen phonon approach for both, the
tetragonal and orthorhombic distorted crystal cells. For the BCT
cell, we find reasonable agreement with experiments only for the
phonons stemming from vibrations of the Bi, O$_{Cu}$ and O$_{Sr}$
atoms, which are, however, still around 15\% lower in frequency
than their experimental counterparts. We predict the opposite
attribution for the two phonons, which have been associated to
vibrations of the Sr and Cu atoms, respectively by experiments on
Y- or Pb-substituted samples. Even more severe is the disagreement
of the O$_{Bi}$ mode, where we predict a value that is only half
of what is experimentally attributed. Due to the distortion along
one of the BiO bond directions, giving rise to an orthorhombic
cell, more vibrations become Raman-active with a total of 12 A$_g$
modes. There are two main effects induced by this distortion. The
first is a coupling of the $c$-axis vibration of O$_{Bi}$ with the
in-plane vibrations in the SrO plane. Secondly, several $a$-axis
vibrations appear due to the symmetry lowering. Among them, the
one associated to the O$_{Bi}$ atoms exhibits a frequency of
424$^{-1}$. It is very near to a peak observed in the measured
Raman spectra which is generally assigned as the $c$-axis O$_{Bi}$
vibration. Our results demonstrate clearly that the distortions
have a huge effect, and should be taken into account when a study
of lattice vibration of Bi2212 is attempted. In spite of this, the
final assignment of the phonon frequencies is not yet fully
solved, and calculations of Raman spectra under different
scattering geometries might be of help in the future.

\begin{acknowledgments}
We are grateful to E. Ya. Sherman for fruitful discussions on
phonon frequencies. This work was partly funded by MIUR-`Progetto
Giovani Ricercatori', and benefitted from support by the Austrian
Science Fund (FWF, project P13430) and the EU RTN network
EXCITING, contract HPRN-2002-00317. The computer facilities were
granted by an INFM project ``Iniziativa Trasversale Calcolo
Parallelo'' at the CINECA supercomputing center.
\end{acknowledgments}


\begin{figure}
 \begin{center}
  \caption{\label{fig1}(Color online) In (a) half of the primitive unit of the body
  centered tetragonal (BCT) cell with space group $I4/mmm$ of
  Bi-2212 is depicted.  In (b) and (c) the BiO plane is sketched  in
  absence or presence of the distortion, respectively. The
  orthorhombic cell is rotated by 45$^{\circ}$ in the $ab$ plane
  with respect to tetragonal one.}
 \end{center}
\end{figure}

\begin{figure*}
 \begin{center}
  \caption{\label{fig2}(Color online) Fermi surface cross cuts in the k$_x$-k$_y$ plane
  (k$_z$ = 0) for the BCT cell with (a) experimental and (b) optimized
  lattice constants and atomic positions (see Tab.~\ref{tab1}). (c)
  The same as in (b),  but in the presence of a BiO terminated (001)
  surface,  where the atomic positions were further relaxed in order
  to account for the breaking of the bonds at the surface layer.}
 \end{center}
\end{figure*}

\begin{figure}
 \begin{center}
  \caption{\label{fig3}Band structure of the undistorted (a) and
  distorted (b) orthorhombic cell along the $\Gamma-M$ direction in
  the orthorhombic Brillouin zone; the tetragonal \textit{\=M} point
  lies in the middle of the $\Gamma-M$ line.}
 \end{center}
\end{figure}

\begin{figure}
 \begin{center}
  \caption{\label{fig4}(Color online) Fermi surface cross cuts in the k$_x$-k$_y$ plane
  (k$_z$ = 0) for the AD-ORTH structure: the rotated square (dashed line)
  whose corners  are the $\Gamma$ point in the middle of the two axis
  compares with the tetragonal Brillouin zones shown in Fig. \ref{fig2}}
 \end{center}
\end{figure}

\begin{table}
 \begin{center}
  \caption{\label{tab1}Lattice parameters and atomic coordinates
  (in units of lattice constants) for the BCT structure of Bi-2212:
  columns 1 to 3 refer to different optimization levels
  (see text for the details), while in the last column
  the experimental data taken from Ref.\onlinecite{sunshine1988} are listed.}
  \renewcommand{\arraystretch}{1.2}
  \begin{ruledtabular}
   \begin{tabular}{ccccc}
    & Full-Opt
    & Vol-Opt
    & Atom-Opt
    & Exp \\ \hline
    Vol [\AA$^3$]
    & 202.2
    & 202.2
    & 226.3
    & 226.3 \\
    c/a
    & 8.258
    & 8.065
    & 8.065
    & 8.065 \\
    $\Delta(E)[\text{mRy}]$
    & -67.4
    & -65.1
    & 0
    & - \\
    z$_{\text{Cu}}$
    & 0.0499
    & 0.0498
    & 0.0491
    & 0.0543 \\
    z$_{\text{Sr}}$
    & 0.1100
    & 0.1103
    & 0.1058
    & 0.1091 \\
    z$_{\text{Bi}}$
    & 0.2008
    & 0.2015
    & 0.1949
    & 0.1989 \\
    z$_{\text{O}_{Cu}}$
    & 0.0514
    & 0.0513
    & 0.0503
    & 0.0510 \\
    z$_{\text{O}_{Sr}}$
    & 0.1331
    & 0.1328
    & 0.1296
    & 0.1200 \\
    z$_{\text{O}_{Bi}}$
    & 0.1991
    & 0.2000
    & 0.1929
    & 0.1980 \\
   \end{tabular}
  \end{ruledtabular}
 \end{center}
\end{table}

\begin{table}
 \begin{center}
  \caption{\label{tab2}Frequencies $\omega$ (in cm$^{-1}$) and
  eigenvectors of the 6 $A_{1g}$  c-axis modes in the BCT cell.}
  \renewcommand{\arraystretch}{1.2}
  \begin{ruledtabular}
   \begin{tabular}{rrrrrrr}
    \multicolumn{1}{c}{$\omega$} &
    \multicolumn{1}{c}{Cu}&
    \multicolumn{1}{c}{Sr}&
    \multicolumn{1}{c}{Bi}&
    \multicolumn{1}{c}{O$_{Cu}$}&
    \multicolumn{1}{c}{O$_{Sr}$}&
    \multicolumn{1}{c}{O$_{Bi}$}\\ \hline
    576 &   0.03 &   0.02 & --0.27 & --0.19 &   \textbf{0.94} &   0.09 \\
    356 &   0.05 &   0.27 &   0.05 & \textbf{--0.94} & --0.19 &   0.03 \\
    233 &   0.00 & --0.03 &   0.21 & --0.06 &   0.14 & \textbf{--0.96} \\
    163 & --0.33 &   \textbf{0.87} & --0.27 &   0.22 & --0.03 & --0.10 \\
    116 & \textbf{--0.92} & --0.23 &   0.25 & --0.12 &   0.07 &   0.08 \\
    50  & --0.17 & --0.34 & \textbf{--0.86} & --0.11 & --0.24 & --0.21 \\
   \end{tabular}
  \end{ruledtabular}
 \end{center}
\end{table}

\begin{table}
 \begin{center}
  \caption{\label{tab3}Lattice parameters and atomic coordinates
  (in unit of lattice constants) for the AD-ORTH structure of Bi-2212:
  the experimental data are taken from Ref.\onlinecite{miles1998}}
  \renewcommand{\arraystretch}{1.2}
  \begin{ruledtabular}
   \begin{tabular}{ccccc}
    & Full-Opt
    & Vol-Opt
    & Atom-Opt
    & Exp \\ \hline
    Vol [\AA$^3$]
    & 428.1
    & 428.1
    & 446.7
    & 446.7\\
    c/a
    & 5.827
    & 5.711
    & 5.711
    & 5.711 \\
    b/a
    & 0.963
    & $\simeq 1$
    & $\simeq 1$
    & $\simeq 1$ \\
    $\Delta(E)[\text{mRy}]$
    & -52.4
    & -35.2
    & 0
    & - \\
    z$_{\text{Cu}}$
    & 0.2019
    & 0.2013
    & 0.2011
    & 0.1978 \\
    z$_{\text{Sr}}$
    & 0.1441
    & 0.1435
    & 0.1445
    & 0.1405 \\
    z$_{\text{Bi}}$
    & 0.0543
    & 0.0528
    & 0.0542
    & 0.0525 \\
    z$_{\text{O}^{1}_{Cu}}$
    & 0.1995
    & 0.1991
    & 0.1991
    & 0.1976 \\
    z$_{\text{O}^{2}_{Cu}}$
    & 0.2006
    & 0.2002
    & 0.2002
    & 0.1994 \\
    z$_{\text{O}_{Sr}}$
    & 0.1192
    & 0.1192
    & 0.1193
    & 0.1189 \\
    z$_{\text{O}_{Bi}}$
    & 0.0568
    & 0.0554
    & 0.0571
    & 0.0500 \\
    x$_{\text{Cu}}$
    & 0.2497
    & 0.2497
    & 0.2496
    & 0.2506 \\
    x$_{\text{Sr}}$
    & 0.2526
    & 0.2524
    & 0.2536
    & 0.2500 \\
    x$_{\text{Bi}}$
    & 0.2269
    & 0.2255
    & 0.2202
    & 0.2258 \\
    x$_{\text{O}^{1}_{Cu}}$\footnotemark[1]
    & 0.0000
    & 0.0000
    & 0.0000
    & 0.0012 \\
    x$_{\text{O}^{2}_{Cu}}$\footnotemark[1]
    & 0.5000
    & 0.5000
    & 0.5000
    & 0.4993\\
    x$_{\text{O}_{Sr}}$
    & 0.2666
    & 0.2663
    & 0.2675
    & 0.2750 \\
    x$_{\text{O}_{Bi}}$
    & 0.1705
    & 0.1726
    & 0.1703
    & 0.1561
   \end{tabular}
  \end{ruledtabular}
  \footnotetext[1]{The O$_{Cu}$ atoms retain their tetragonal positions
  in our calculations.}
 \end{center}
\end{table}

\begin{table*}
 \begin{center}
  \caption{\label{tab4}Frequencies $\omega$ (in cm$^{-1}$) and
  eigenvectors of the 12 $A_{g}$ modes in the AD-ORTH cell.
  The atoms vibrate along the c- and a- axis, as indicated.}
  \renewcommand{\arraystretch}{1.2}
  \begin{ruledtabular}
   \begin{tabular}{rrrrrrrrrrrrr}
    \multicolumn{1}{c}{$\omega$} &
    \multicolumn{1}{c}{Cu}&
    \multicolumn{1}{c}{Cu}&
    \multicolumn{1}{c}{Sr}&
    \multicolumn{1}{c}{Sr}&
    \multicolumn{1}{c}{Bi}&
    \multicolumn{1}{c}{Bi}&
    \multicolumn{1}{c}{O$^1_{Cu}$}&
    \multicolumn{1}{c}{O$^2_{Cu}$}&
    \multicolumn{1}{c}{O$_{Sr}$}&
    \multicolumn{1}{c}{O$_{Sr}$}&
    \multicolumn{1}{c}{O$_{Bi}$}&
    \multicolumn{1}{c}{O$_{Bi}$} \\
    &
    \multicolumn{1}{c}{c-axis}&
    \multicolumn{1}{c}{a-axis}&
    \multicolumn{1}{c}{c-axis}&
    \multicolumn{1}{c}{a-axis}&
    \multicolumn{1}{c}{c-axis}&
    \multicolumn{1}{c}{a-axis}&
    \multicolumn{1}{c}{c-axis}&
    \multicolumn{1}{c}{c-axis}&
    \multicolumn{1}{c}{c-axis}&
    \multicolumn{1}{c}{a-axis}&
    \multicolumn{1}{c}{c-axis}&
    \multicolumn{1}{c}{a-axis}  \\ \hline
    597 & 0.02 & 0.00 & 0.02 & 0.00 & --0.25 & --0.02 &
    --0.09 & --0.10 & \textbf{0.95} & 0.12 & 0.05 & 0.04 \\
    424 &  0.00 & 0.00 & --0.03 & 0.01 & 0.00 & --0.21 &
    0.04 & 0.03 & 0.05 & --0.06 & --0.01 & \textbf{--0.97} \\
    367 & --0.09 & --0.28 & --0.19 & --0.03 & --0.02 & --0.01 &
    \textbf{0.67} & \textbf{0.63} & 0.12 & 0.08 & --0.08 & 0.06 \\
    356 & 0.02 & \textbf{--0.95} & 0.06 & --0.10 & 0.01 & 0.00 &
    --0.12 & --0.27 & --0.04 & 0.03 & 0.01 & --0.02 \\
    289 & --0.01 & --0.06 & 0.02 & 0.09 & --0.22 & 0.16 &
    0.02 & 0.12 & 0.00 & \textbf{--0.65} & \textbf{0.69} & 0.00 \\
    210 & --0.02 & 0.10 & 0.00 & 0.09 & --0.09 & 0.00 &
    \textbf{0.65} & \textbf{--0.62} & --0.08 & 0.26 & 0.31 & --0.02 \\
    192 & --0.03 & 0.04 & 0.21 & \textbf{--0.45} & --0.22 & --0.14 &
    --0.19 & 0.27 & --0.15 & \textbf{0.55} & \textbf{0.49} & --0.03 \\
    163 & --0.30 & --0.01 & \textbf{0.88} & 0.17 & --0.17 & 0.06 &
    0.13 & 0.07 & --0.01 & --0.07 & --0.20 & --0.03 \\
    132 & --0.07 & 0.10 & 0.06 & \textbf{--0.86} & 0.08 & 0.15 &
    0.18 & --0.17 & 0.08 & --0.33 & --0.19 & --0.01 \\
    106 & --0.41 & 0.01 & --0.07 & --0.04 & 0.02 & \textbf{--0.86} &
    --0.01 & --0.08 & 0.01 & --0.20 & 0.03 & 0.20 \\
    104 & \textbf{0.84} & 0.00 & 0.22 & --0.04 & --0.22 & --0.37 &
    0.14 & 0.03 & --0.05 & --0.14 & --0.11 & 0.09 \\
    47 & 0.16 & 0.00 & 0.28 & 0.01 & \textbf{0.86} & --0.08 &
    0.06 & 0.07 & 0.21 & 0.02 & 0.29 & 0.02 \\
   \end{tabular}
  \end{ruledtabular}
 \end{center}
\end{table*}

\end{document}